\documentclass[aps,prl,reprint,longbibliography,superscriptaddress]{revtex4-1}
\usepackage{amsmath}    
\usepackage{graphicx}   
\usepackage{verbatim}   
\usepackage{color}      
\usepackage{subfigure}  
\usepackage{hyperref}   
\usepackage[table]{xcolor}
\usepackage{multirow}
\usepackage{hhline}
\usepackage{booktabs}
\usepackage{makecell}
\usepackage{gensymb}

\definecolor{lightgray}{gray}{0.9}

\newcommand{\er}{Er$^{3+}$}

\newcommand{\moo}{MoO$_3$}
\newcommand{\ptosto}{PbTiO$_{3}$:SrTiO$_{3}$}
\newcommand{\tio}{TiO$_2$}
\newcommand{\pbwo}{PbWO$_4$}
\newcommand{\cawo}{CaWO$_4$}

\newcommand{\zy}[2]{$Z_{#1}\xrightarrow[]{}Y_{#2}$}

\newcommand{\dc}{$\degree{}$C}

\newcommand{\wn}{cm$^{-1}$}

\makeatletter
\newcommand{\balancecolsandclearpage}{
  \close@column@grid
  \cleardoublepage
  \twocolumngrid
}

\begin{document}
\title{Erbium-Implanted Materials for Quantum Communication Applications}

\author{Paul Stevenson}
\affiliation{Department of Physics, Northeastern University, Boston, Massachusetts 02115, USA}

\author{Christopher M Phenicie}
\affiliation{Department of Electrical Engineering, Princeton University, Princeton, New Jersey 08544, USA}

\author{Isaiah Gray}
\affiliation{Department of Electrical Engineering, Princeton University, Princeton, New Jersey 08544, USA}

\author{Sebastian P Horvath}
\affiliation{Department of Electrical Engineering, Princeton University, Princeton, New Jersey 08544, USA}

\author{Sacha Welinski}
\affiliation{Thales Research and Technology, 1 Avenue Augustin Fresnel, 91767 Palaiseau, France}

\author{Austin M Ferrenti}
\affiliation{Department of Chemistry, Princeton University, Princeton, New Jersey 08544, USA}

\author{Alban Ferrier}
\affiliation{Chimie ParisTech, PSL University, CNRS, Institut de Recherche de Chimie Paris, Paris, France, 75005}
\affiliation{Faculté des Sciences et Ingénierie,  Sorbonne Université, UFR 933, 75005 Paris, France}

\author{Philippe Goldner}
\affiliation{Chimie ParisTech, PSL University, CNRS, Institut de Recherche de Chimie Paris, Paris, France, 75005}

\author{Sujit Das}
\affiliation{Department of Materials Science and Engineering, University of California, Berkeley, Berkeley, 94720, California, USA}
\affiliation{Department of Physics, University of California, Berkeley, 94720, California, USA}

\author{Ramamoorthy Ramesh}
\affiliation{Department of Materials Science and Engineering, University of California, Berkeley, Berkeley, 94720, California, USA}
\affiliation{Materials Sciences Division, Lawrence Berkeley National Laboratory, Berkeley, 94720, California, USA}
\affiliation{Department of Physics, University of California, Berkeley, 94720, California, USA}

\author{Robert J Cava}
\affiliation{Department of Chemistry, Princeton University, Princeton, New Jersey 08544, USA}

\author{Nathalie P de Leon}
\affiliation{Department of Electrical Engineering, Princeton University, Princeton, New Jersey 08544, USA}

\author{Jeff D Thompson}
\email{jdthompson@princeton.edu}
\affiliation{Department of Electrical Engineering, Princeton University, Princeton, New Jersey 08544, USA}

\date{\today}
\begin{abstract}
Erbium-doped materials can serve as spin-photon interfaces with optical transitions in the telecom C-band, making them an exciting class of materials for long-distance quantum communication. However, the spin and optical coherence times of \er{} ions are limited by currently available host materials, motivating the development of new \er{}-containing materials. Here, we demonstrate the use of ion implantation to efficiently screen prospective host candidates, and show that disorder introduced by ion implantation can be mitigated through post-implantation thermal processing to achieve inhomogeneous linewidths comparable to bulk linewidths in as-grown samples. We present optical spectroscopy data for each host material, which allows us to determine the level structure of each site, allowing us to compare the environments of \er{} introduced via implantation and via doping during growth. We demonstrate that implantation can generate a range of local environments for \er{}, including those observed in bulk-doped materials, and that the populations of these sites can be controlled with thermal processing.
\end{abstract}
\maketitle

\section{Introduction}

Quantum networks have myriad applications, including fundamentally secure communication\cite{Ekert1991}, modular quantum computing\cite{Wehner2018}, and precision measurement\cite{Komar2014}. This has motivated substantial theoretical and experimental efforts to realize scalable schemes for both long distance quantum communication and microwave-to-optical transduction. A major goal is to develop single atom quantum memories that can be deployed in quantum repeater architectures\cite{Kimble2008,Atature2018a}.

Erbium-based systems are promising candidates for both single atom quantum memories in long distance quantum networks and microwave-to-optical quantum state transduction. Several desirable properties have been demonstrated in various \er{}-containing systems, such as long electron\cite{Dantec2021} and nuclear\cite{Rancic2018} spin coherence times, and telecom-wavelength spin-photon interfaces\cite{Raha2020}. Realizing these properties simultaneously in a single system, however, is an area of great current interest.

Many of the systems characterized for quantum information applications are yttrium-based crystals (\textit{e.g.} Y$_2$SiO$_5$, YVO$_4$). \er{} readily substitutes for the Y$^{3+}$ site, but the large spin bath formed by the $^{89}$Y nuclear spins has deleterious effects on spin and optical coherence times. While extremely long optical coherence times can be observed in \er{}:Y$_2$SiO$_5$ ($T_2$=4.38\,ms at 1.5\,K and 7\,T\cite{Bottger2009}), these optical coherence times are ultimately limited by the presence of $^{89}$Y nuclei \cite{Bottger2006}, motivating the development of nuclear spin-free host materials. 

However, \er{} in non-yttrium materials has been studied in a variety of other contexts, such as infrared-to-visible upconversion\cite{Golesorkhi2020}, laser gain media\cite{Sun2002}, and as probes of the local crystal environment\cite{Baker1976}, demonstrating that there is a rich chemical space still to explore for quantum information applications. Efficiently exploring this space and screening potential new materials remains an outstanding challenge\cite{Ferrenti2020}.

A further challenge is encountered when integrating the current generation of \er{}-containing materials with nanophotonic structures; while nanophotonic integration can enable single-ion detection\cite{Dibos2018} and quantum state transduction\cite{Bartholomew2020}, the emitters addressed by the device are typically within hundreds of nanometers of the surface. Surface-related noise processes (trapped charges, dangling bonds \textit{etc}) can substantially degrade the optical and spin coherence of nearby emitters\cite{Dibos2018,Raha2020,Sangtawesin2018a1}. Developing new material systems with reduced sensitivity to surface noise --- either through non-polar local symmetry, or fortuitously small Stark coefficients --- may help enable stable, narrow optical transitions in these nanophotonic devices. These new materials would enable nanophotonic integration with implanted single crystal substrates as a ``top-down'' alternative to other cavity-emitter architectures such as nanoparticles in fiber cavities\cite{Casabone2021}.

We recently demonstrated that \er{} implanted in rutile \tio{} exhibits inhomogeneous optical linewidths below 500 MHz, among the narrowest inhomogeneous linewidths reported for any host material for \er{}\cite{Phenicie2019a} and especially remarkable for ion-implanted \er{}. Here we build on this result, using the same methodology to explore the properties of several other Er-implanted materials. By utilizing ion implantation to introduce \er{} into a variety of materials, we are able to rapidly generate and screen new material systems while also ensuring our results are representative of the properties of the near-surface ions required for nanophotonic integration.

The paper is structured as follows: in Section I, we outline the criteria for host material selection; Section II briefly outlines the experimental methods; Sections III-V present results for cubic, non-polar, and polar symmetry crystals, respectively; and Section VI concludes with general observations.

\section{I: Host Material Criteria}

\begin{table*}
\begin{tabular}{|c||c|c|c||c|c||c|c|c|c|}
\hline
 & \multicolumn{3}{c||}{MgO} & \multicolumn{2}{c||}{\ptosto{}} & \multicolumn{4}{c|}{ZnS} \\ \hhline{|=|===|==|====|}
 & \multicolumn{2}{c|}{Experiment} & \multicolumn{1}{c||}{Cubic Model \cite{Borg1970}} & \multicolumn{2}{c||}{Experiment} & \multicolumn{2}{c|}{Site 1} & \multicolumn{2}{c|}{Site 2} \\ \hline
\zy{1}{1} \wn{} & \multicolumn{2}{c|}{6491.5} & \multicolumn{1}{c||}{N/A} & \multicolumn{2}{c||}{6511.7}  & \multicolumn{2}{c|}{6489.3} & \multicolumn{2}{c|}{6486.7} \\ \hline
 $n$ (energies in \wn{})& Z$_n$ & Y$_n$ & Z$_n$ &  Z$_n$ & Y$_n$ & Z$_n$ & Y$_n$ & Z$_n$ & Y$_n$ \\ \hline
 1 & 0.0, 0.08 & 0.0 & 0.0, 0.0  & 0.0 & 0.0 &  0.0 & 0.0 & 0.0 & 0.0 \\ \hline
 2 & 110.4 & 48.3, 48.6 & 111.2  & 14.21 & 73.7 & 7.4 & 6.7 & 11.2 & 5.7 \\ \hline
 3 & 134.8 & 98.7 & 139.7  & 113.47 & 75.1 & 25.6 & 30.7 & 31.4 & 16.7 \\ \hline
 4 & * & * & 619.2  & * & 204.8 & * & 39.6 & * & 49.0 \\ \hhline{|=|==|=|==|==|==|}
  Lifetime / ms& \multicolumn{2}{c|}{20.1 (19.6)} & \multicolumn{1}{c||}{N/A} & \multicolumn{2}{c||}{3.9 (6.3)} & \multicolumn{2}{c|}{4.6 (8.4)} & \multicolumn{2}{c|}{5.6 (8.4)} \\ \hline
\end{tabular}
\caption{Crystal field energies and other properties of cubic-symmetry host materials. Lifetimes predicted for purely magnetic-dipole transitions are given in parenthesis. Crystal field levels for the $Y$ levels are given relative to the \zy{1}{1} transition energy. Levels assigned to strain-split quartet states are given in the same table entry. States which are not observed in our experiments are denoted with asterisks.}
\label{Tab:cubic}
\end{table*}

To narrow our search, we begin with a coarse set of criteria for prospective host crystals which can be determined from the elemental composition or from readily available reference data. Specifically, we search for material properties that allow for long \er{} coherence times, and a material bandgap sufficient to host the optical transition. An extensive tabulation of materials satisfying these properties was presented in Reference \cite{Ferrenti2020}.

Long spin coherence times require an environment with low magnetic noise. Thus, a good host material should be diamagnetic, composed of elements with nuclear spin-free isotopes, and have minimal paramagnetic impurities. Spin coherence is ultimately limited by spin-lattice relaxation, but this can typically be suppressed by going to sufficiently low temperatures\cite{Orbach1961}.

Coherent optical transitions are an essential element of a spin-photon interface. Optical coherence introduces two requirements; first, the band gap of the host material must be large enough such that emission from the defect optical transition is not re-absorbed by the host material ($>0.83$\,eV for \er{}), and second, the optical transition frequency should be stable in the presence of the solid state environment. One approach to realizing this second requirement is to optimize crystal growth and processing to remove sources of noise\cite{Chu2014}, while a complimentary strategy is to minimize the sensitivity of the qubit to these noise sources. If the symmetry of the crystal site does not allow a permanent electric dipole moment, then the Stark shift from electric field noise is quadratic instead of linear. Narrow ($<$ 500MHz) inhomogeneous linewidths previously reported in Er-implanted \tio{} were attributed to the non-polar symmetry (and thus vanishing first-order Stark shift) of the \er{} site in this material\cite{Phenicie2019a}, and inversion-symmetric group IV-vacancy defects in diamond have significantly less spectral diffusion near surfaces and in nanostructures\cite{Sipahigil2014,Rose2018a} compared to NV centers\cite{Chu2014}.

Below, we present the optical characterization of a selection of Er-implanted materials which meet the criteria outlined earlier: they are composed of elements which have stable nuclear spin-free isotopes, bandgaps greater than 1\,eV, and which span a wide range of potential coordination environments for implanted \er{}. By probing both the inhomogeneous linewidths and implantation yield for particular sites we are able to gain insight into both the static disorder in the crystal and the sensitivity of \er{} to noise sources in the crystal. These crystals span a wide range of potential \er{} coordination environments: cubic (MgO, \ptosto{}, ZnS), non-cubic and non-polar (\tio{}, PbWO$_4$), and polar symmetries (\moo{}, ZnO).

This wide-ranging screening approach allows us to identify MgO, \pbwo{}, \tio{} and ZnO as particularly promising candidates, each showing narrow inhomogeneous linewidths for near-surface ions. In several material systems we also demonstrate the importance of post-implantation thermal processing.

\section{II: Experimental Approach}

\er{} is incorporated into host materials \textit{via} ion implantation (Innovion) at energies up to 350\,keV, which corresponds to an implantation depth of $\approx$ 100\,nm in the materials studied here (see Supplementary Information for full implantation recipe). The implantation process creates a highly non-equilibrium distribution of \er{} coordination environments and implantation-induced vacancies. To identify different \er{} environments in the host crystals and characterize their properties, we employ photoluminescence-excitation (PLE) spectroscopy, described below. These techniques enable us to ascertain the site-specific level structure of different \er{} environments and their associated inhomogeneous linewidths. Samples are mounted in a sample-in-vapor optical cryostat held at 4.5\,K unless otherwise noted. Multiple samples are mounted on the same copper cold finger with GE varnish, using copper barriers between samples to prevent scattered light from exciting fluorescence in adjacent samples. 

An achromatic doublet lens outside the cryostat is used to focus excitation light and collect emission from the sample, and a tunable narrowband laser (CTL 1500, Toptica) is used to excite the sample. The resulting emission spectrum is measured using a grating spectrometer with a cooled InGaAs detector array (PyLon-IR, Teledyne Princeton Instruments), enabling us to generate two-dimensional maps of the emission spectrum for various excitation wavelengths. Fast thermalization of the \er{} excited states (relative to the millisecond radiative lifetime\cite{Liu2005}) results in identical emission spectra for each \er{} site regardless of which transition was excited; thus when \er{} can occupy multiple different sites the emission spectrum can be used to group excitation peaks originating from \er{} ions occupying the same site.

With site-specific excitation and emission frequencies, we can extract the ground and excited state energies for each site. The ground state splits into 8 states $Z_1-Z_8$, (5 in cubic symmetry), while the excited state splits into 7 states $Y_1-Y_7$ (5 in cubic symmetry). For each site we also measure the excited state lifetime, which we can compare to the expected magnetic dipole (MD) decay rate\cite{Dodson2012}. This allows us to estimate the role of forced electric dipole moments in the decay rate, which are only allowed in sites without inversion symmetry.

\section{III: Cubic Systems}

\subsection{MgO}

\textbf{Structure \& Properties:} MgO adopts the cubic halite structure, with octahedrally-coordinated (local $O_h$ symmetry) Mg$^{2+}$ and O$^{2-}$. The only native nuclear spin present at an appreciable level is $^{25}$Mg ($I=\frac{5}{2}$, 10\% natural abundance). 

MgO has a relatively low refractive index ($n=1.71$)\cite{Stephens1952}, which is advantageous for improving refractive index contrast in hybrid nanophotonic devices. The large band gap of MgO (7.8\,eV) is sufficient to host not only telecom-wavelegth optical transitions, but can also host defects with shorter wavelength transitions.

The samples used in this work had a (001) surface and were purchased from MTI Corp. Samples were stored in a dessicator when not in use because of the hygroscopic nature of MgO. Prior to implantation, no \er{} emission was observed.

\begin{figure}
  \centering
  \includegraphics[]{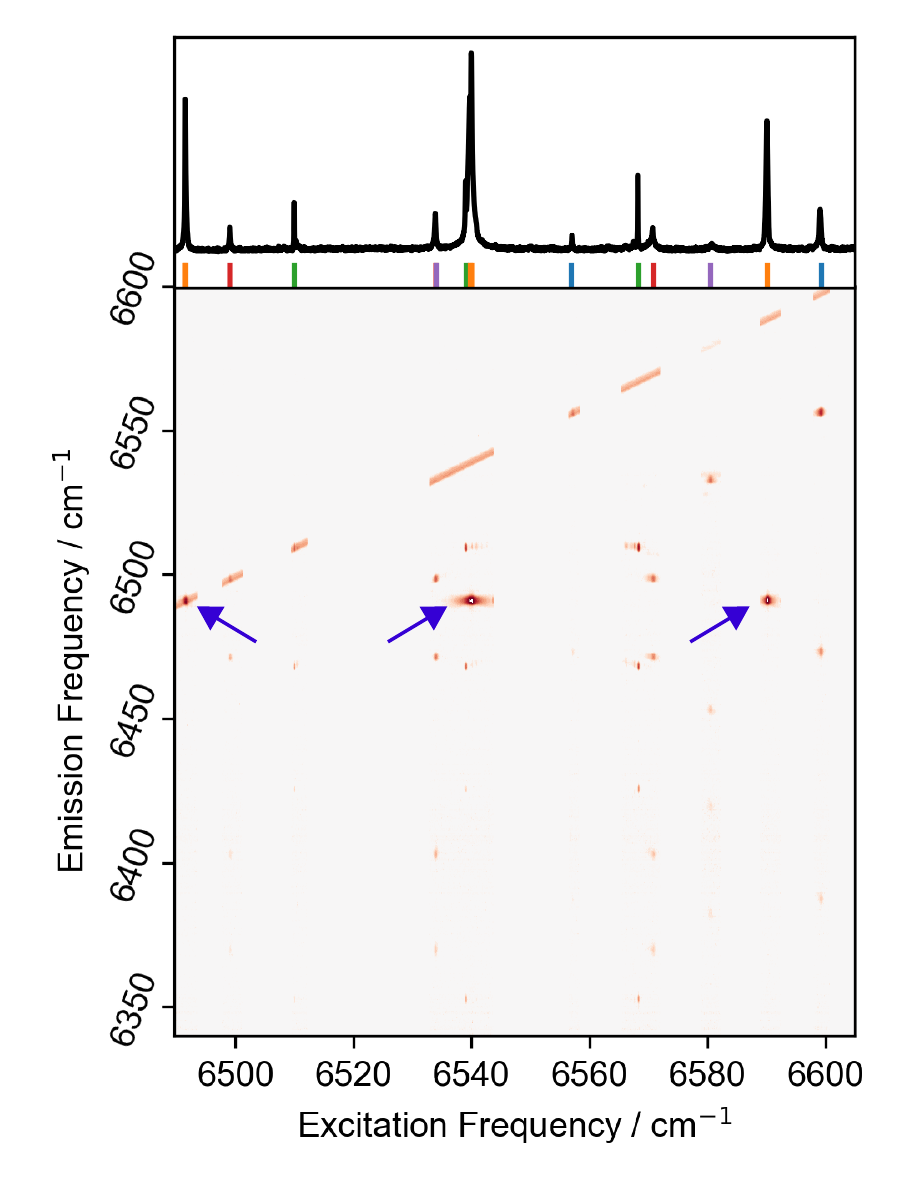}
  \caption{Top panel: excitation spectrum of annealed \er{}:MgO at T=4.5\,K, showing all transitions observed. Colored bars indicate different sites. Bottom panel: excitation-emission spectrum of the same sample, plotted on a log intensity axis to show all peaks observed. The diagonal feature at $\omega_{exc}=\omega_{em}$ is from scattered excitation light, and provides a wavelength reference for the spectrometer.} 
  \label{Fig:MgO_2D}
\end{figure}

\begin{figure}
  \centering
  \includegraphics[]{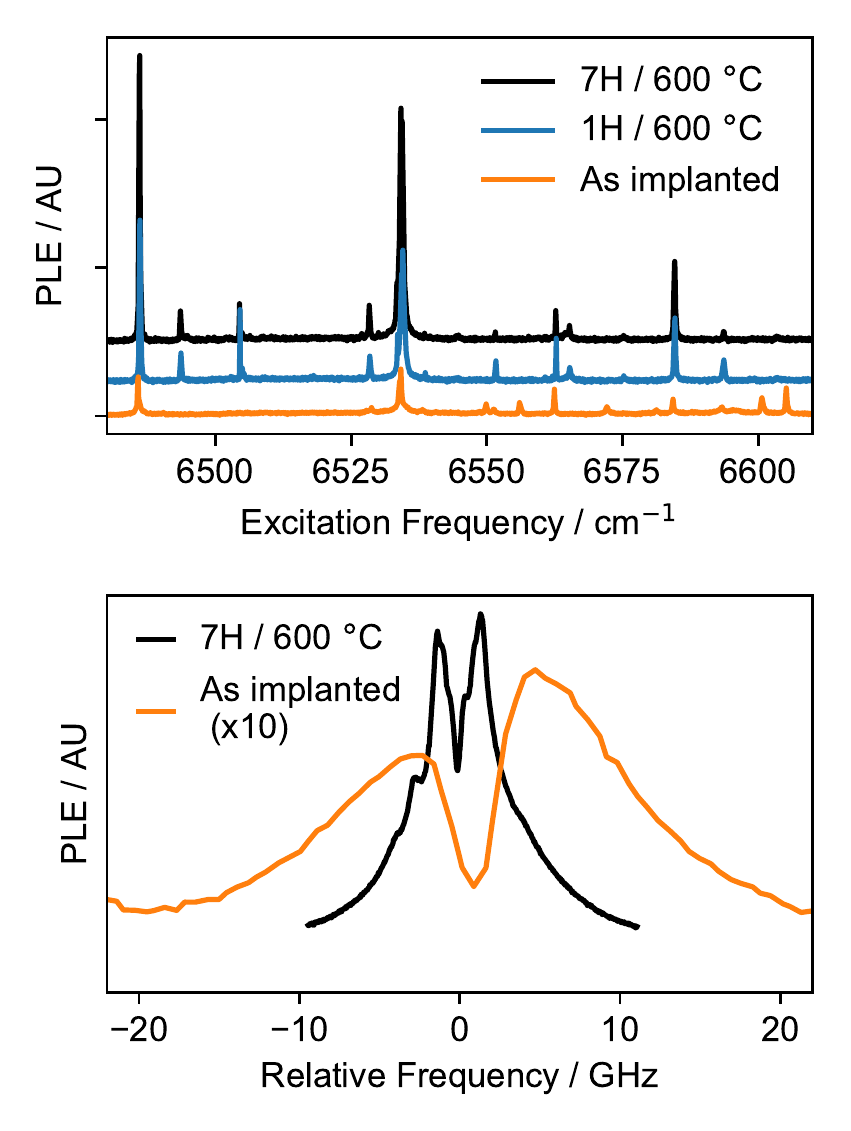}
  \caption{Upper panel: Changes in the number and intensities of peaks in MgO as a function of annealing conditions, described in the main text. Bottom panel: High-resolution scan of the \zy{1}{1} transition, showing the change in linewidth on annealing.} 
  \label{Fig:MgO_anneal}
\end{figure}

\textbf{Optical Spectrum:} The excitation-emission spectrum of MgO shown in Figure \ref{Fig:MgO_2D} has a rich structure with peaks corresponding to at least five distinct incorporation sites. We focus our discussion on the most intense set of peaks, highlighted with blue arrows in Fig. \ref{Fig:MgO_2D} (other energy levels are tabulated in the Supplementary Information). The lowest two excitation peaks show small splittings (0.08\,\wn{}, 0.3\,\wn{} respectively) which are suggestive of some degeneracy-lifting perturbation. Substantial unresolved fine structure is evident in the \zy{1}{1} peak (Fig. \ref{Fig:MgO_anneal}), making quantitative determination of inhomogeneous linewidth challenging; however, we can put an upper bound of 2\,GHz on the width of each peak (post thermal processing), comparable to the inhomogeneous linewidth observed in many Yttrium-based crystals\cite{Liu2005}.

The fluorescence lifetime of the $Y_1$ level is 20.1\,ms (see Supplementary InformationI). We can estimate the expected radiative lifetime of a magnetic dipole (MD) transition by scaling the lifetime in the free \er{} ion (98.3\,ms) by $n^3$ (to account for the higher density of states in the crystal), which yields an estimated MD lifetime of 19.6\,ms\cite{Dodson2012}. The good agreement with the experimental value suggests that there is no forced electric dipole contribution and that non-radiative relaxation pathways are not significant in this case. 

Previous ESR measurements have reported that \er{} occupies the octahedral Mg$^{2+}$ substitutional site\cite{Baker1976,Borg1970}, and has a quartet ground state ($\Gamma_8$ in the Koster notation), rather than the usual Kramers doublet. Strain can lift this fourfold degeneracy\cite{Baker1976}, consistent with the closely-spaced sets of peaks observed in our measurements. Using previously-reported crystal field parameters\cite{Borg1970,Lea1962}, we determine that the transition from $Y_1$ to the lowest non-quartet state ($\Gamma_7\xrightarrow[]{}\Gamma_6$) is symmetry-forbidden. To determine the ground-state splittings, we increase the temperature of the sample to populate the $Y_2$ level. Emission from this level allows us to determine the ground state splittings, which are in excellent agreement with previously reported models based on ESR measurements for the substitutional Mg$^{2+}$ site (Table \ref{Tab:cubic})\cite{Borg1970}. 

\textbf{Annealing:} \er{}-implanted MgO was annealed in air for a total of seven hours at 600\dc{}. Previous studies of MgO implanted with Fe have indicated aggregation and precipitation above this temperature\cite{White1989}. Post-anneal spectra show a seven-fold increase in the intensity of the $O_h$ site. We also observe a number of peaks after annealing which we suggest may be associated with structural configurations where charge compensation occurs locally (as in Nd$^{3+}$ in CaF$_2$\cite{Kiss1963}), though the detailed structure and symmetry of this site are unknown. (The level structure for this site is detailed in the Supplementary Information). A number of small peaks initially present in the as-implanted sample disappear on annealing. Figure \ref{Fig:MgO_anneal} shows the dependence of the peak intensities on annealing conditions.

The increase in intensity of the Site 1 peaks with annealing indicates that thermal processing is important for \er{}-implanted MgO, though the results here likely do not represent the optimal annealing method. The detailed mechanism, and optimization, of this thermal processing will be the focus of future work.

\subsection{\ptosto}
\textbf{Structure \& Properties}: PbTiO$_3$ and SrTiO$_3$ are both perovskite structures, with cuboctahedral coordination of the A-site (Pb$^{2+}$, Sr$^{2+}$) and octahedral coordination of the B-site (Ti$^{4+}$). Below 105\,K, SrTiO$_3$ undergoes a symmetry-lowering tetragonal distortion\cite{Cowley1996}, lowering the local symmetry of the A- and B-sites to $D_{2d}$ and $C_{4h}$ respectively\cite{Evarestov2011}. Both Ti and Pb have isotopes with nuclear spin, $^{207}\mathrm{Pb}$ (I$=\frac{1}{2}$, 22\% natural abundance), $^{47}\mathrm{Ti}$ (I$=\frac{5}{2}$, 7\% natural abundance), and $^{49}\mathrm{Ti}$ (I$=\frac{7}{2}$, 5\% natural abundance).

This sample is not implanted with \er{}; doping is instead introduced during growth of the PbTiO$_3$ layer. 100\,nm of PbTiO$_3$ is overgrown on a SrTiO$_3$ substrate \textit{via} pulsed-laser deposition, as detailed in the Supplementary Information. The \er{}-doping is modulated such that only one unit-cell layer is doped with \er{} at 0.1 mol$\%$, then ten layers are undoped, then another unit-cell layer is doped, and so on. Both SrTiO$_3$ and PbTiO$_3$ have high refractive indices ($n=2.28$\cite{Bond1965} and $n<2.5$\cite{Singh1972} respectively).

\begin{figure}
  \centering
  \includegraphics[]{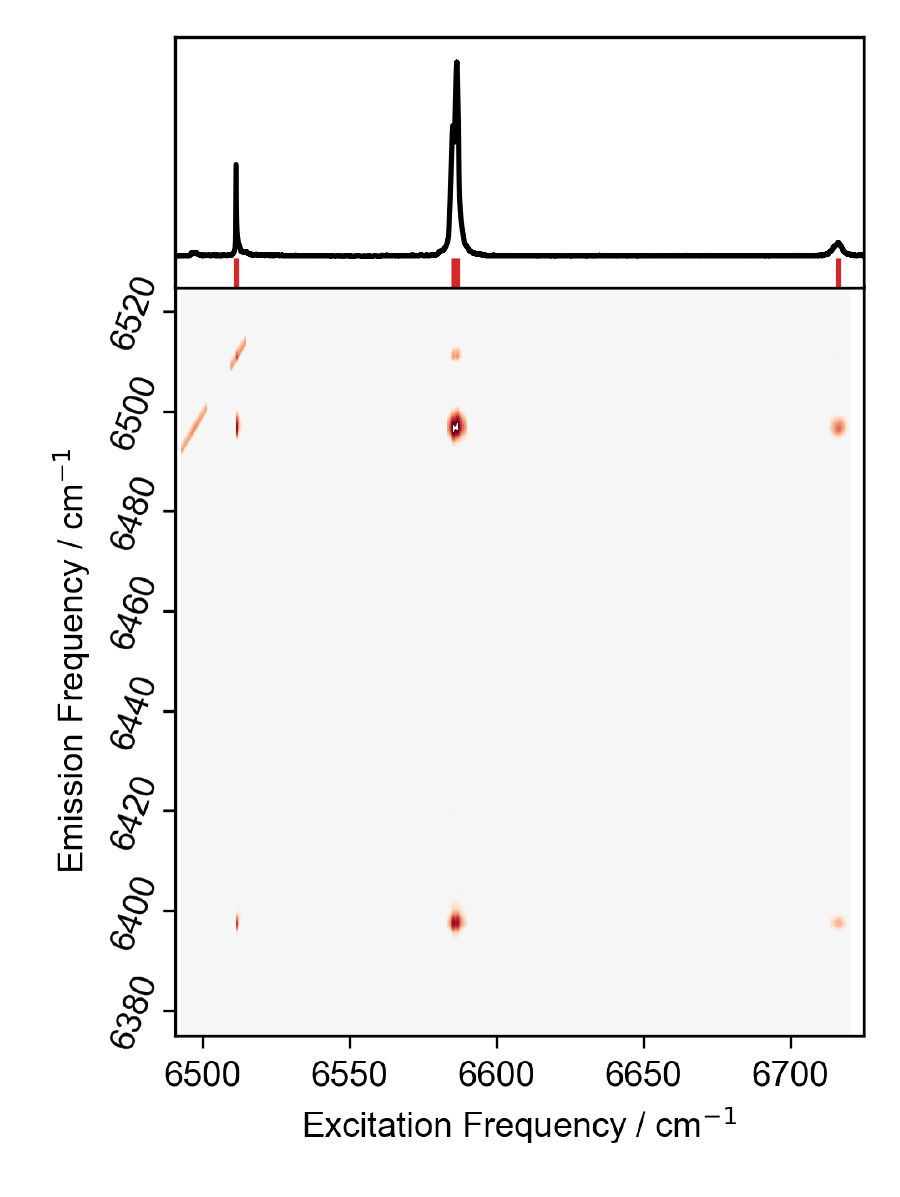}
  \caption{Top panel: Excitation spectrum of \er{}-doped \ptosto{}, showing all transitions observed. Bottom panel: Excitation-emission spectrum of the same sample, plotted on a log intensity axis to show all peaks observed.} 
  \label{Fig:pto_2D}
\end{figure}

\textbf{Optical Spectrum:}
In contrast to MgO, \er{}:\ptosto{} shows a much simpler spectrum, shown in Figure \ref{Fig:pto_2D}. All peaks show the same emission pattern, consistent with a single occupied site. The extracted crystal field levels are given in Table \ref{Tab:cubic}.

The inhomogeneous linewidth of the \zy{1}{1} transition is $\approx$9\,GHz, with an asymmetric tail. The excited state lifetime of the $Y_1$ state is 3.9\,ms. The expected magnetic dipole radiative lifetime is 6.3\,ms, which is indicative of a forced electric dipole contribution to the emission.

While our measurements do not allow us to unambiguously assign a site to the incorporated \er{}, we note that ESR measurements of \er{} in BaTiO$_3$ have suggested \er{} can substitute in either the A- or B-site\cite{Dunbar2004}, and the better matching of ionic radius between \er{} and Pb$^{2+}$ with respect to Ba$^{2+}$ suggests A-site (distorted cubeoctahedral) substitution might be expected. The A-site ($D_{2d}$ symmetry at low temperature) permits an electric dipole contribution to the radiative lifetime \textit{via} $d$-$f$ mixing, while the B-site symmetry ($C_{4h}$) does not.

\subsection{ZnS}
\textbf{Structure \& Properties:} ZnS adopts the zinc blende structure, composed of tetrahedrally coordinated Zn$^{2+}$ and S$^{2-}$. Only the $^{67}$Zn isotope has non-zero spin ($I=\frac{5}{2}$, natural abundance 4\%). ZnS has a refractive index $n=2.27$\cite{Debenham1984} and a band gap of 3.5\,eV.

The samples described in this work were purchased from SurfaceNet, with (111) orientation, grown by a seeded vapor-phase free growth method. No \er{} emission was observed prior to implantation.

\begin{figure}
  \centering
  \includegraphics[]{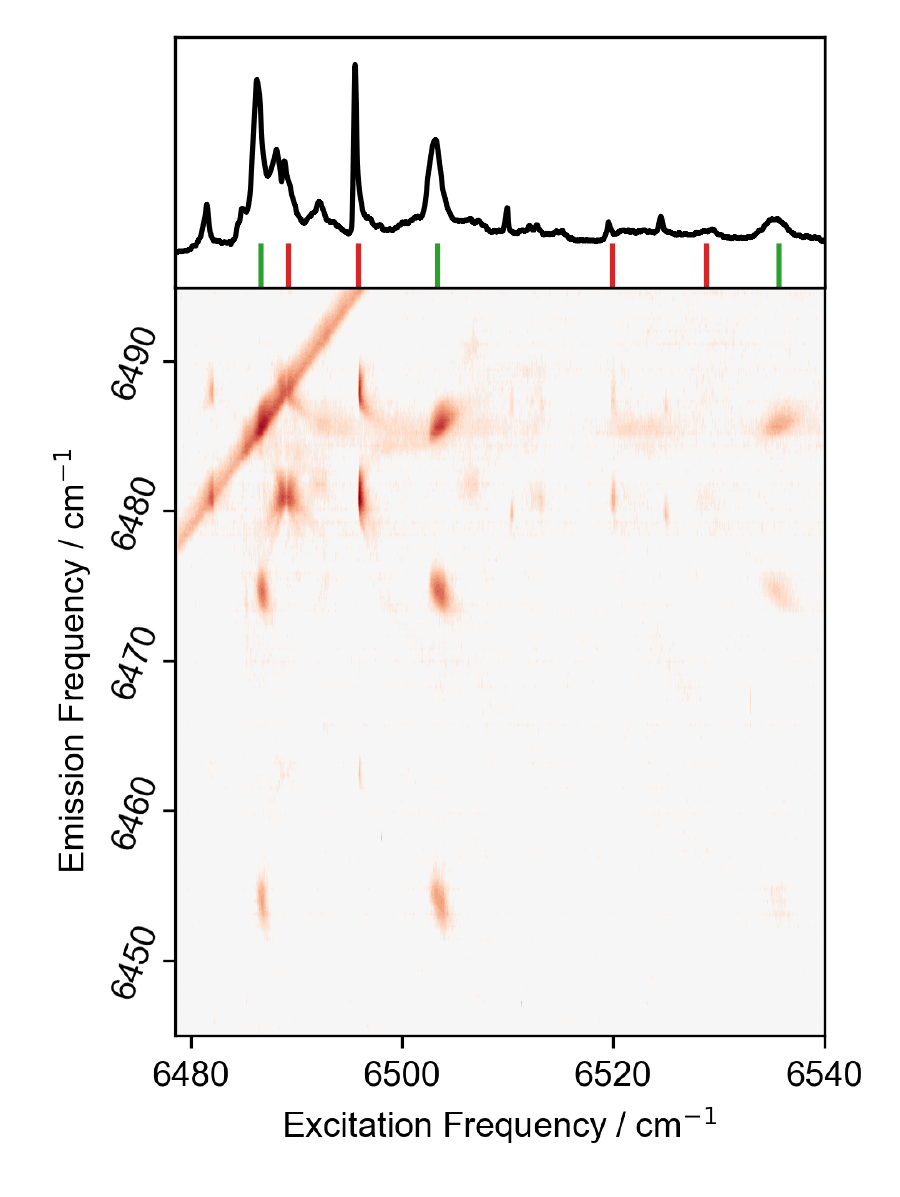}
  \caption{Top panel: Excitation spectrum of \er{}-implanted ZnS (annealed), showing all transitions observed. Bottom panel: Excitation-emission spectrum of the same sample, plotted on a log intensity axis to show all peaks observed.} 
  \label{Fig:zns_2D}
\end{figure}

\textbf{Optical Spectrum:} Two distinct sites (Sites 1 and 2) can be identified in the ZnS excitation-emission spectrum, shown in Figure \ref{Fig:zns_2D}. We note that for both sites, the transitions span a much smaller energy range than either MgO or \ptosto{}. Extracted crystal field levels are reported in Table \ref{Tab:cubic}.

The measured excited state lifetimes for both Sites 1 (4.6\,ms) and 2 (5.6\,ms) are shorter than expected for a pure MD transition (8.4\,ms). The congested spectrum makes accurate determination of the inhomogeneous linewidths difficult, even with site-selective measurements, but both Sites 1 and 2 have \zy{1}{1} inhomogeneous linewidths of $>$25\,GHz.

Emission in the visible wavelength range from \er{}-implanted ZnS has been previously reported using cathodoluminescence\cite{Yu1979}, where two sites were also observed. The $Z$ levels we determine for Site 2 are in excellent agreement with the reported levels for the substitutional Zn$^{2+}$ site. Moreover, the population of this site increased with annealing in a manner consistent with previous reports\cite{Yu1979}.

However, the parameters we determine for Site 1 are not consistent with the reported levels for the interstitial site identified in Reference \cite{Yu1979} despite similar implantation fluences and subsequent annealing. One possible explanation for this discrepancy is the excitation method; cathodoluminescence and resonant excitation may excite sites with different efficiencies. Though we do not propose a detailed structural assignment for Site 1, we note that some charge compensation must occur in ZnS with \er{} defects (as with all the materials we present here); differences in charge compensation (\textit{e.g.} local or remote, compensating ion identity, local configuration) would be expected to give rise to distinct sites. 

\textbf{Annealing:} \er{}-implanted ZnS was annealed in air at 350\dc{} for 3 hours, as described in Reference \cite{Yu1979}. Prior to annealing, Site 1 was observable, but no clear emission from Site 2 was detected. After annealing, emission from Site 2 increased substantially, having similar maximum peak height to Site 1. 

\section{IV: Non-Cubic, Non-Polar Systems}

\begin{table}
\begin{tabular}{|c||c|c||c|c|}
\hline
 & \multicolumn{2}{c||}{\pbwo{}} & \multicolumn{2}{c|}{\tio{}} \\ \hhline{|=|==|==|}
 \zy{1}{1} / \wn{} & \multicolumn{2}{c||}{6517.7} & \multicolumn{2}{c|}{6575.7} \\ \hline
$n$ (energies in \wn{})& Z$_n$ & Y$_n$ & Z$_n$ & Y$_n$ \\ \hline
1 & 0.0 & 0.0 & 0.0 & 0.0 \\ \hline
2 & 10.7 & 3.6 & 35.0 & 14.8 \\ \hline
3 & 31.9 & 47.6 & 40.8 & 22.9 \\ \hline
4 & 51.4 & 103.4 & 119.6 & 96.3 \\ \hline
5 & 189.5 & 134.9 & 169.7 & * \\ \hline
6 & 228.7 & 147.6 & 208.3 & * \\\hline
7 & 253.3 & 165.4 & 269.8 & * \\ \hline
8 & 282.1 & N/A & * & N/A \\ \hhline{|=|==|==|}
Lifetime / ms & \multicolumn{2}{c||}{7.7 ($>9.1$)} & \multicolumn{2}{c|}{5.25 (5.2-6)\cite{Phenicie2019a}} \\ \hline
\end{tabular}
\caption{Crystal field energies and other properties of non-polar symmetry host materials. Lifetimes predicted for purely magnetic-dipole transitions are given in parenthesis. Crystal field levels for the $Y$ levels are given relative to the \zy{1}{1} transition energy}
\label{Tab:nonpolar}
\end{table}

\subsection{\pbwo}
\textbf{Structure \& Properties:} PbWO$_4$ adopts the Scheelite crystal structure, in which the optical properties of doped rare-earth ions have been extensively studied\cite{Enrique1971}. Both Pb and W have isotopes with nuclear spin, $^{207}\mathrm{Pb}$ (I$=\frac{1}{2}$, 22\% natural abundance) and $^{183}\mathrm{W}$ (I$=\frac{5}{2}$, 14\% natural abundance). We observe \er{} emission in our PbWO$_4$ prior to implantation, suggesting low levels of native doping; this would suggest that extra care may be required during synthesis to minimize additional paramagnetic defects. Similar background levels of \er{} have been observed in other nominally lanthanide-free materials\cite{Wolfowicz2021a}.

\pbwo{}, like all Scheelite structures, is birefringent. The refractive indices at 1530\,nm are $n_o<2.21$ and $n_e<2.16$\cite{Baccaro1997}. Numerous values have been reported for the band gap of \pbwo{}, but all are $>$3.5\,eV. \cawo{}, which we compare \pbwo{} with below, has the same structure and similar optical properties ($n_o=1.88$, $n_e=1.9$\cite{Bond1965}, band gap $4.2$\,eV). In \er{}-doped \cawo{}, \er{} is known to occupy the  Ca$^{2+}$ site, which has local $S_4$ symmetry\cite{Enrique1971,Cornacchia2007}.

The \pbwo{} substrate with (001) orientation was purchased from MTI Corp. The comparison \cawo{} sample was grown \textit{via} the Czochralski method, with \er{} doping of approximately 0.8\,ppb\cite{BertainPC}.

\textbf{Optical Spectrum:} 
We compare the optical properties of three samples - \er{}-implanted \pbwo{}, \pbwo{} with native \er{} (presumably introduced during crystal growth), and as a reference sample, \er{}-doped \cawo{}. By comparing the intensities of the observed spectra in the implanted and unimplanted \pbwo{}, we estimate the background \er{} doping levels to be on the $<$ppb-level.

\begin{figure}
  \centering
  \includegraphics[]{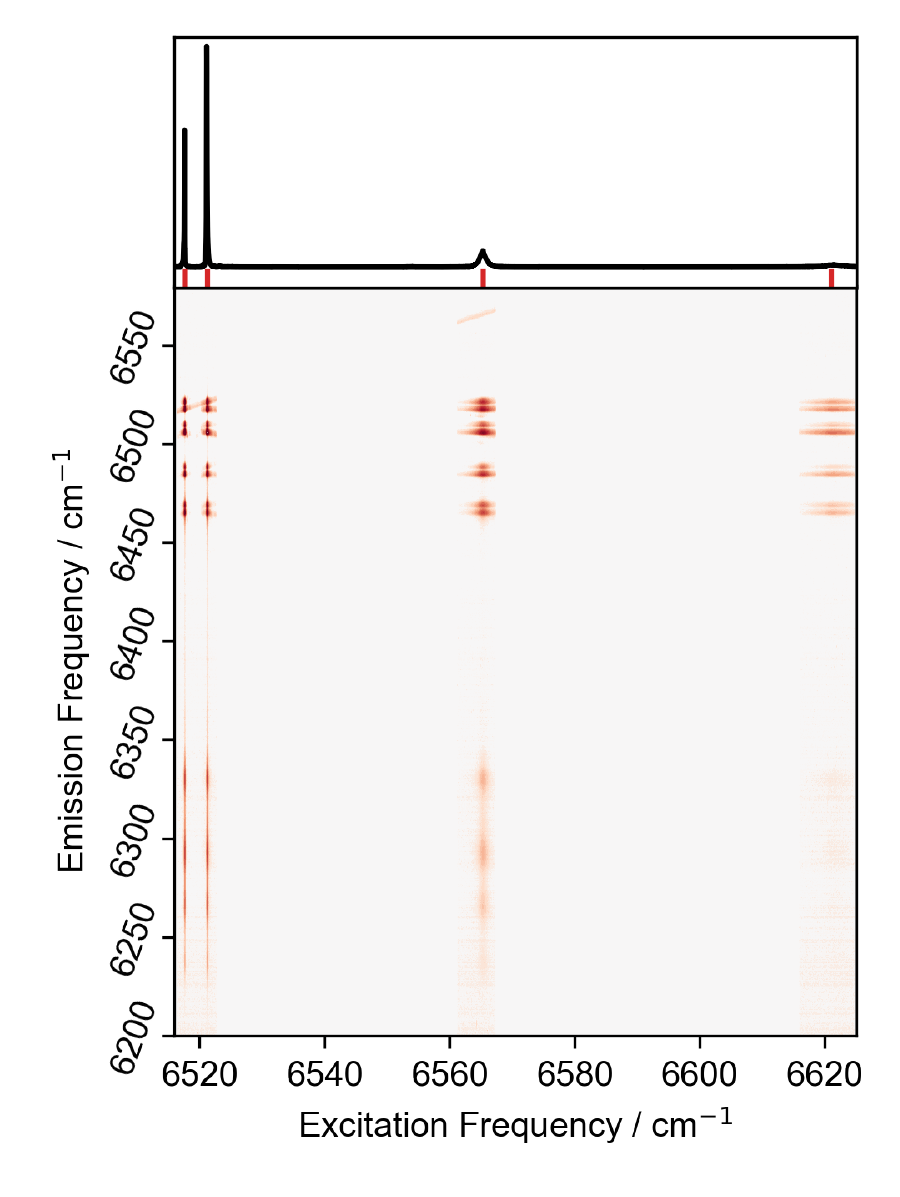}
  \caption{Top panel: Excitation spectrum of \er{}-implanted \pbwo{} (annealed). Bottom panel: Excitation-emission spectrum of the same sample, plotted on a log intensity axis to show all peaks observed.} 
  \label{Fig:pbo4_2D}
\end{figure}

The excitation-emission spectrum of implanted \pbwo{} (Figure \ref{Fig:pbo4_2D}) shows only one site. The peaks are all clearly resolvable, allowing the determination of all eight $Z$ levels and all seven $Y$ levels (Table \ref{Tab:nonpolar}). The splittings reported are qualitatively similar to those known for isostructural \er{}-doped \cawo{}\cite{Enrique1971}. 

\begin{figure}
  \centering
  \includegraphics[]{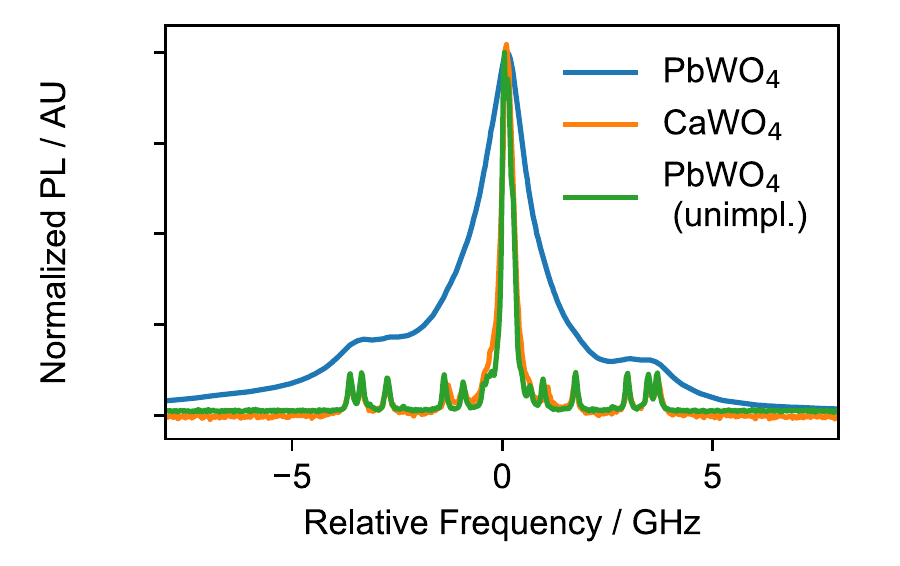}
  \caption{Top panel: high resolution scan of \zy{1}{1} transition for implanted \pbwo{}, bulk-doped CaWO$_4$, and unimplanted \pbwo{} with trace \er{} impurities. The same hyperfine structure is observed in all cases. \pbwo{} and \cawo{} spectra are shifted by their respective \zy{1}{1} transition energies, but the frequency axis is not otherwise altered.} 
  \label{Fig:pbo4_lines}
\end{figure}

The excited state lifetime of implanted \pbwo{} is 7.7\,ms, shorter than the $9.3$\,ms expected from a pure MD transition, suggesting a forced electric dipole contribution. The \zy{1}{1} transition of the implanted sample (Figure \ref{Fig:pbo4_lines}) shows significant structure, but the FWHM of the central peak is approximately 1\,GHz. As noted earlier, our \pbwo{} samples have trace \er{} contamination pre-implantation, enabling us to directly compare implanted and bulk \er{} in \pbwo{} (Figure \ref{Fig:pbo4_lines}). We find that the peak positions are identical between the implanted and trace-\er{} samples, though the bulk \er{} inhomogeneous linewidths are threefold narrower. This might suggest there is still some residual disorder from implantation; however, we also note that our local \er{} concentration is high ($\approx$100\,ppm) in the implanted region, and 1\,GHz inhomogeneous linewidths have been observed in bulk-doped \er{}:\cawo{} of similar concentrations\cite{Sun2002}.

To further explore the similarities between \cawo{} and \pbwo{}, we compare the the spectrum of the implanted \pbwo{} and a bulk-doped \cawo{} sample. The same hyperfine structure around the \zy{1}{1} transitions is observed for both \cawo{} and \pbwo{}, which arises from the $23\%$ natural abundance $^{167}$Er. The near perfect correspondence between the \pbwo{} and \cawo{} spectra suggest their hyperfine constants (and thus underlying $g$-tensors) are extremely similar, from which we conclude that \er{} occupies a similar site in both cases. In \cawo{}, \er{} is known to occupy the substitutional cation site\cite{Enrique1971,Cornacchia2007}.

The prominent hyperfine-split transitions are a result of the large transverse component of the hyperfine tensor\cite{Enrique1971}; this acts in the same manner as a large off-axis magnetic field, giving rise to optical transitions which flip the electron or nuclear spin. Well-resolved hyperfine transitions may find use in optical manipulation of the nuclear spin\cite{Welinski2020}, potentially serving as an ancilla qubit.

\textbf{Annealing:} Implanted \pbwo{} was annealed for one hour at 900\dc{}. Prior to annealing, the signal observed (both magnitude and linewidth) was similar to the native \er{} in the unimplanted crystal, suggesting implanted \er{} is not optically active. After annealing the sample the signal increased by a factor of $>200$.

\subsection{\tio}
\textbf{Structure \& Properties:} \tio{} adopts the rutile structure in our samples. Previous studies of Er-implanted \tio{} demonstrated that the \er{} occupies the substitutional Ti site, with local $D_{2h}$ symmetry\cite{Phenicie2019a}.

Ti has two isotopes with nuclear spin, $^{47}\mathrm{Ti}$ (I$=\frac{5}{2}$, 7\% natural abundance) and $^{49}\mathrm{Ti}$ (I$=\frac{7}{2}$, 5\% natural abundance). \tio{} has a large refractive index ($n>2.4$) and is highly birefringent. The bandgap of \tio{} is 3.0\,eV. The preparation of the samples described below have been previously described\cite{Phenicie2019a}.

\begin{figure}
  \centering
  \includegraphics[]{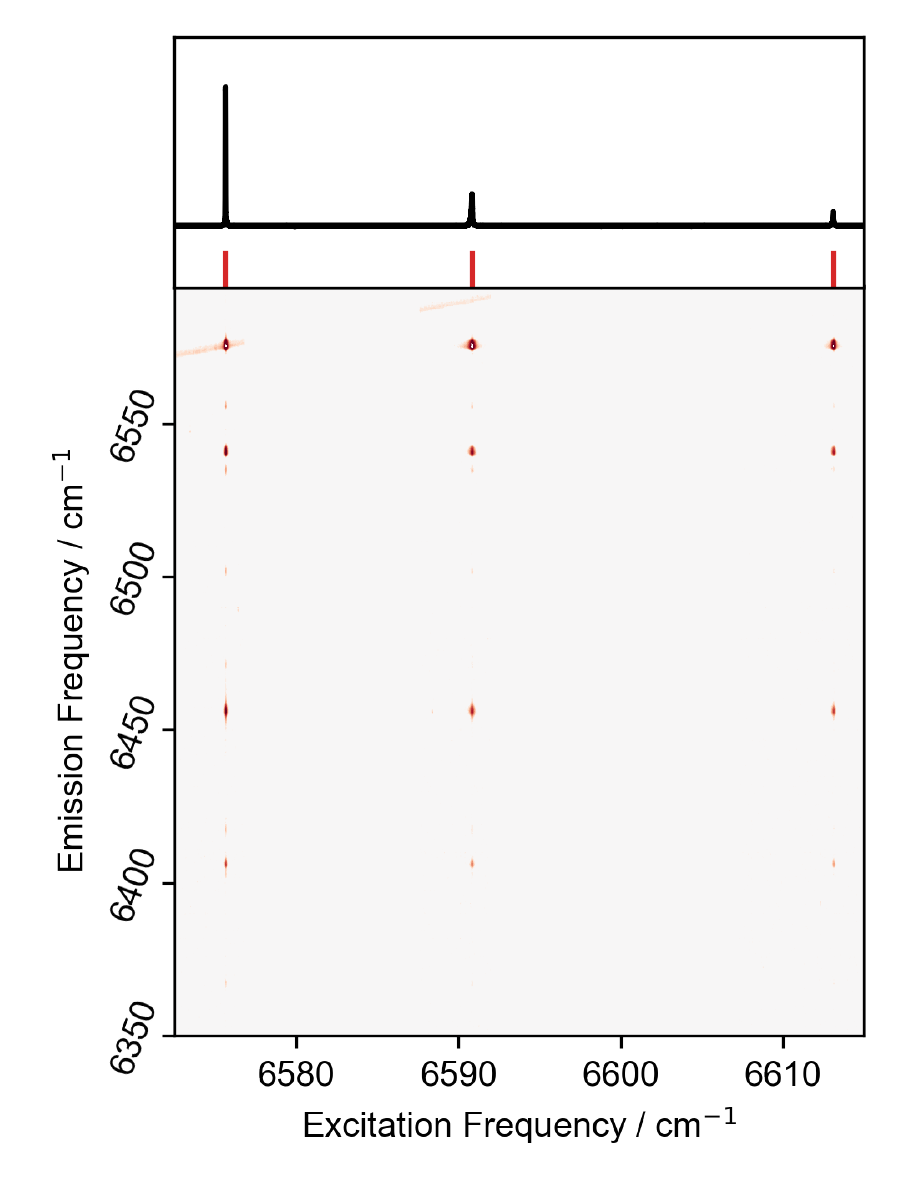}
  \caption{Top panel: Excitation spectrum of \er{}-implanted \tio{} (annealed), showing all transitions observed. Bottom panel:  Excitation-emission spectrum of the same sample, plotted on a log intensity axis to show all peaks observed.} 
  \label{Fig:tio2}
\end{figure}

\textbf{Optical Spectrum:} The excitation spectrum and inhomogeneous linewidth of implanted \tio{} have been previously reported in Ref \cite{Phenicie2019a}. Here, an improved optical setup enables us to report a more complete excitation-emission spectrum (Figure \ref{Fig:tio2}), which is consistent with the previously reported crystal field levels (Table \ref{Tab:nonpolar}). Experimental improvements in resolution and sensitivity allow us to report here several previously-unreported $Z$ levels.

\section{V: Polar Systems}

\begin{table}
\begin{tabular}{|c||c|c||c|c|}
\hline
 & \multicolumn{2}{c||}{\moo{}} & \multicolumn{2}{c|}{ZnO} \\ \hhline{|=|==|==|}
 \zy{1}{1} / \wn{} & \multicolumn{2}{c||}{6512.5} & \multicolumn{2}{c|}{6504.2} \\ \hline
$n$ (energies in \wn{})& Z$_n$ & Y$_n$ & Z$_n$ & Y$_n$ \\ \hline
1 & 0.0 & 0.0 & 0.0 & 0.0 \\ \hline
2 & 36.5 & 55.6 & 19.8 & 14.8 \\ \hline
3 & 135.6 & 82.5 & 27.8 & 22.9 \\ \hline
4 & 163.0 & 117.2 & 116.9 & * \\ \hline
5 & 177.4 & * & 152.9 & * \\ \hline
6 & * & * & 195.4 & * \\\hline
7 & * & * & 217.8 & * \\ \hhline{|=|==|==|}
 Lifetime / ms & \multicolumn{2}{c||}{5.5 (11.0)} & \multicolumn{2}{c|}{5.8 (13.6)} \\ \hline
\end{tabular}
\caption{Crystal field energies and other properties of polar symmetry host materials. Lifetimes predicted for purely magnetic-dipole transitions are given in parenthesis. Crystal field levels for the $Y$ levels are given relative to the \zy{1}{1} transition energy}
\label{Tab:polar}
\end{table}

\subsection{\moo}
\textbf{Structure:} \moo{} forms orthorhombic crystals composed of distorted octahedra of Mo$^{6+}$ coordinated by O$^{2-}$. Only two sites of non-trivial symmetry are found in this crystal, $C_s$ (Mo and O positions) and $C_i$ (interstitial sites). Mo has two isotopes with nuclear spin, $^{95}\mathrm{Mo}$ (I$=\frac{5}{2}$, 15\% natural abundance) and $^{97}\mathrm{Mo}$ (I$=\frac{5}{2}$, 10\% natural abundance). The refractive index of \moo{} is $n=2.07$\cite{Lajaunie2013}. The bandgap of \moo{} is $3.2$\,eV\cite{Sabhapathi1995}. The samples used in the work below were grown by the vapor transport method. 

\textbf{Optical Spectrum:} The spectrum of implanted \moo{} is consistent with a single dominant environment for \er{}, showing a sparse pattern of excitation and emission peaks. The excited state lifetime of \moo{} is 5.5\,ms, shorter than the 11.0\,ms expected from a pure magnetic dipole transition in this material. The inhomogeneous linewidth of the \zy{1}{1} transition is 90\,GHz, which is the broadest linewidth we report here. There are several possible sources of disorder in this system; not only is \er{} in \moo{} sensitive to implantation-induced damage, but the variety of oxidation states that can be adopted by Mo has the potential to serve as an intrinsic source of inhomogeneity.

\begin{figure}
  \centering
  \includegraphics[]{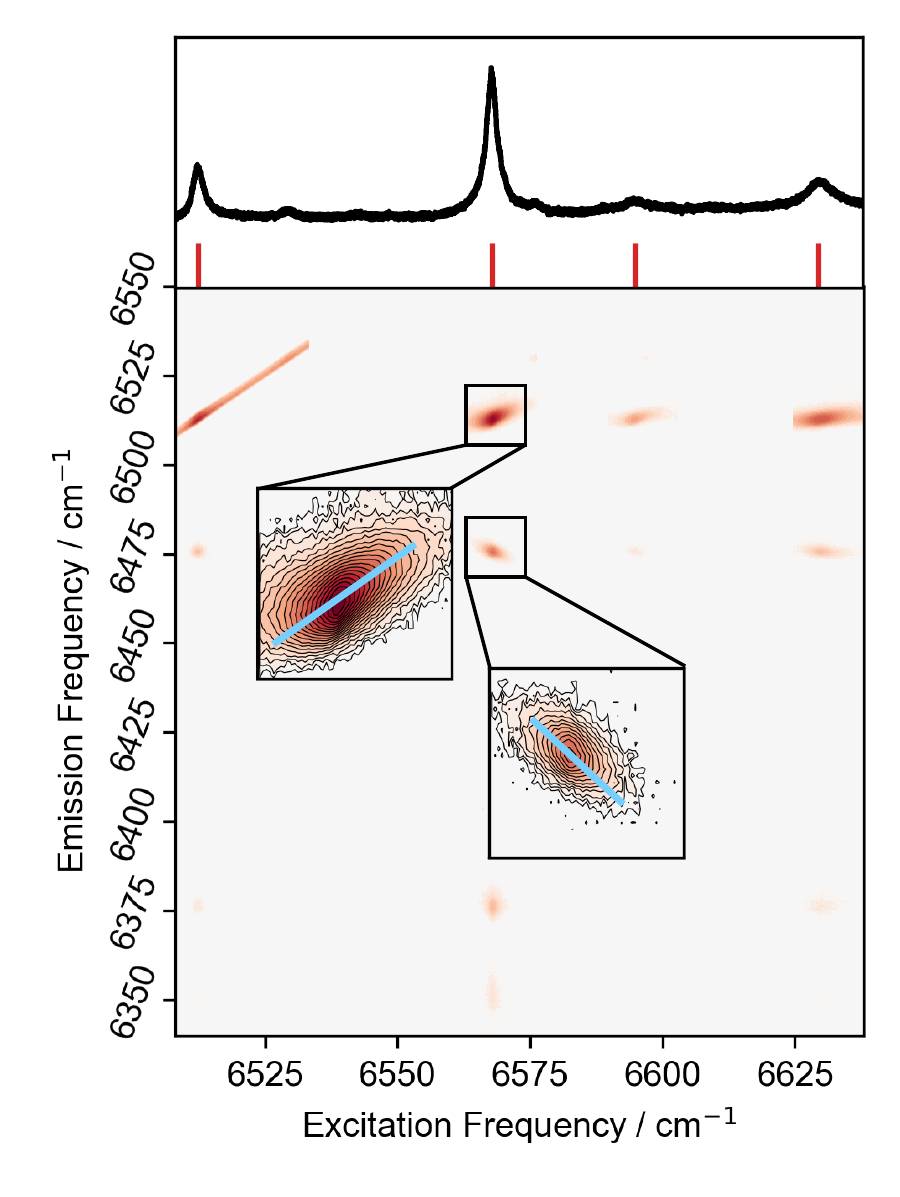}
  \caption{Top panel: excitation spectrum of \er{}-implanted \moo{} (annealed), showing all transitions observed. Bottom panel: excitation-emission spectrum of the same sample, plotted on a log intensity axis to show all peaks observed. Inset shows the correlated lineshape in the excitation-emission spectrum for two peaks.} 
  \label{Fig:moo_2D}
\end{figure}

The broad inhomogeneous linewidths in this sample highlight another source of information accessible from our spectroscopic approach. Several of the two-dimensional lineshapes show distinct slopes (such as the inset of Figure \ref{Fig:moo_2D}) which reveal the correlations between excitation and emission frequencies of sub-ensembles of \er{} ions (\textit{e.g.} a sub-ensemble may have a red-shifted \zy{1}{1} transition, but a blue shifted \zy{1}{2} transition). These correlations describe how the different energy levels of the \er{} ions shift in response to the main source of disorder in the system (electric field, strain, \textit{etc}), which may be useful in site assignment or identifying transitions which are minimally sensitive to the local disorder. This approach is currently limited by our spectrometer resolution ($\approx 1$\,\wn{}), but improvements in detection resolution would allow this information to be extracted for narrower inhomogeneous linewidths.

\textbf{Annealing}: The implanted \moo{} sample was annealed in air at 550\dc{} for 4 hours. Previous studies assessing \er{}-implanted \moo{} as a potential optoelectronic material found this annealing procedure increased the visible emission from \er{} in \moo{}\cite{Vila2014}. \er{} emission was observed both pre- and post-anneal, the main effect of the annealing was the removal of a number of small, barely-resolvable peaks, but did not impact the inhomogeneous linewidth. Because of the irregular size and shape of the crystals, it is not possible to make a quantitative comparison between pre- and post-anneal peak heights. The post-anneal excitation-emission spectrum is consistent with a single environment for the implanted Er.

\subsection{ZnO}
\textbf{Structure:} ZnO adopts the hexagonal Wurtzite structure. The Zn$^{2+}$ ion is coordinated by four O$^{2-}$ in a distorted tetrahedral fashion with $C_{3v}$ symmetry. The polar crystal structure of ZnO gives rise to a piezoelectric effect in the crystal. Only the $^{67}$Zn isotope has non-zero spin ($I=\frac{5}{2}$, natural abundance 4\%).

The refractive index of ZnO is $n=1.93$\cite{Bond1965}, and the bandgap is 3.3\,eV. The samples described here had a (0001) orientation and were purchased from MTI Corp. No emission from \er{} was observed prior to implantation.

\begin{figure}
  \centering
  \includegraphics[]{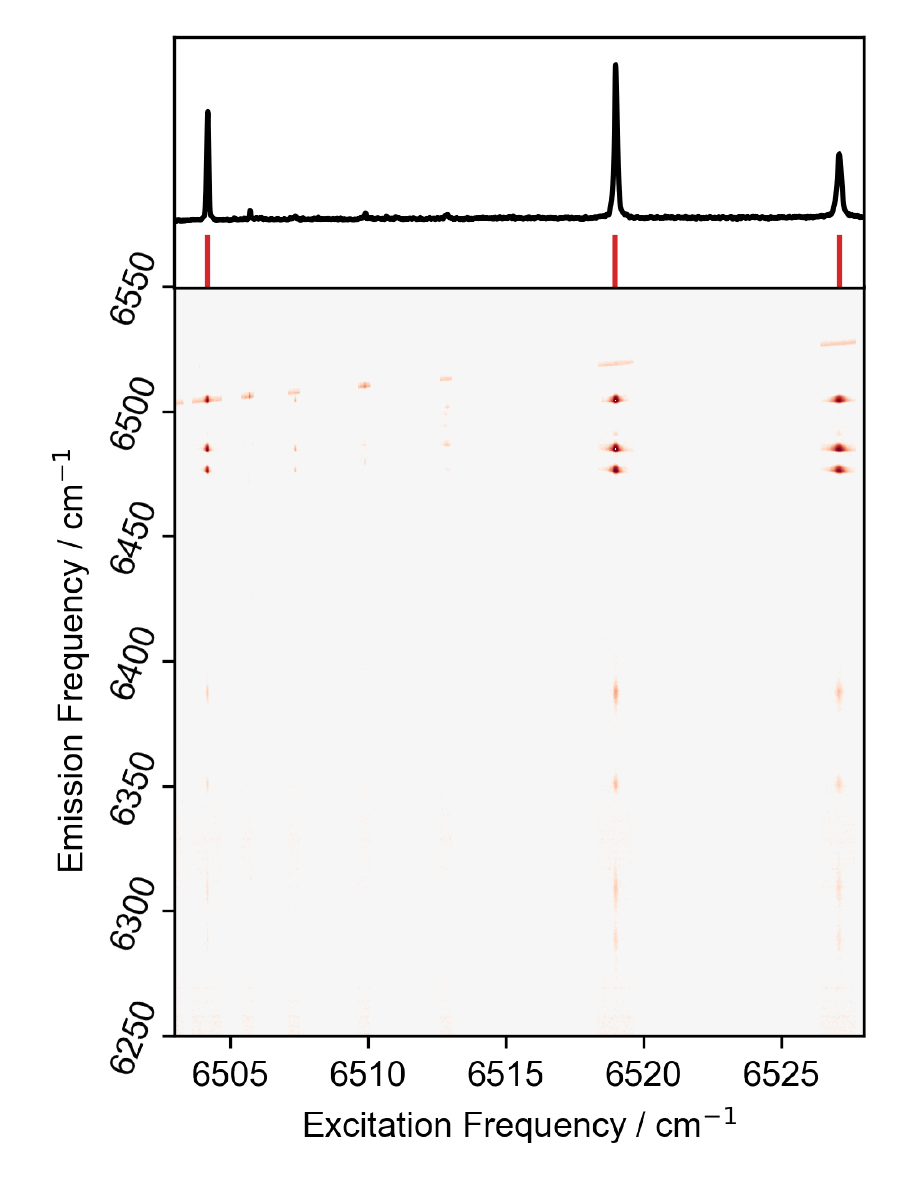}
  \caption{Top panel: Excitation spectrum of \er{}-implanted ZnO (annealed), showing all transitions observed. Bottom panel: Excitation-emission spectrum of the same sample, plotted on a log intensity axis. Several smaller features are visible in the excitation range 6507\wn{} to 6512\wn{}; these are transitions from thermally populated states of the \er{}.} 
  \label{Fig:zno_2D}
\end{figure}

\textbf{Optical Spectrum:} The optical spectrum of Er-implanted ZnO, shown in Figure \ref{Fig:zno_2D}, is dominated by a single site. The inhomogeneous linewidth of the \zy{1}{1} transition is 1.5\,GHz, which is among the narrowest of the materials we present here. This narrow linewidth is observed despite the polar symmetry and piezoelectric nature of the crystal. An interesting point of comparison here are ``Pseudo-$C_{3v}$'' defects in SiC, where similarity between dipoles in the ground and excited states results in minimal sensitivity of the optical transition to electric fields\cite{Udvarhelyi2019}.

The lifetime of \er{}-ZnO is substantially smaller than would be expected for a magnetic-dipole transition (5.8\,ms \textit{vs} 13.6\,ms), which indicates a forced electric dipole contribution.

Rutherford backscattering measurements of implanted ZnO have observed that \er{} occupies the substitutional Zn$^{2+}$ site\cite{Alves2003}, while optical measurements of \er{}-implanted GaN\cite{Alves1999}, which adopts the same structure and has similar lattice constants, also find that \er{} occupies the cation substitutional site. Our observation of a single, well-defined site in the excitation-emission spectrum is consistent with a picture of \er{} substituting for Zn$^{2+}$, where the sparseness of the spectrum suggests either remote charge compensation or one dominant charge compensation configuration.

\textbf{Annealing}: ZnO was annealed at 750\dc{} for a total of sixteen hours; this was the total time at 750\dc{} over three annealing runs of duration 1 hour, 6 hours, and 9 hours. Annealing at higher temperatures is known to lead to out-diffusion of \er{}\cite{Alves2003}. As-implanted, the spectrum showed a wide range of low-intensity, spectrally broad features; after annealing for one hour, a new set of peaks with much narrower inhomogeneous linewidths was observed. The intensity of these peaks increased with longer annealing times, reaching a maximum increase of 10x relative to the as-implanted peak height. Spectra for these annealing steps are shown in the Supplementary Information.

\section{VI: Discussion}

Ion implantation and doping during growth are two distinct approaches to introducing \er{} ions into a host material; understanding to what extent the two approaches can generate similar coordination environments of \er{} is necessary for exploring new host materials for quantum communication applications. We find that, where the coordination environment for the doped samples is known (\textit{i.e.}, MgO, PbWO4 and ZnO), ion implantation is capable of generating the same local environment. 

However, we are also able to identify several other sites occupied by implanted \er{} in various samples using ion implantation. In MgO, we can resolve at least four other sites in addition to the $O_h$ site (detailed in the Supplementary Information); multiple sites are observable in ZnS; and the spectrum of ZnO does not resolve into a clear single site until the sample has been annealed. To explain this observation, we consider the effects of charge compensation and implantation-induced damage to the host crystal. In all systems described here, \er{} is an aliovalent defect, requiring charge compensation in the crystal; different configurations of local charge compensation can give rise to multiple spectrally distinct peaks, as in the case of Nd$^{3+}$ in CaF$_2$\cite{Kiss1963} and \er{}:Si\cite{Przybylinska1996,Gritsch2021}. Additionally, ion implantation generates vacancy defects which may produce additional coordination environments of \er{} which would not be observed in a sample doped during growth. 

The effect of ion implantation damage is not limited to generating new, distinct, \er{} sites, but can also give rise to disorder in the crystal host. The inhomogeneous linewidth reflects the extent of the disorder, as well as the sensitivity of the \er{} optical transitions to this disorder. In multiple implanted systems (MgO, \pbwo{}, \tio{}, ZnO) we see inhomogeneous linewidths on the order of a few GHz (summarized in Table \ref{Tab:linewidths}), comparable to what is typically observed in conventional host materials such as \er{}:Y$_2$SiO$_5$\cite{Liu2005}, suggesting that the effect of static disorder in the implanted samples can be similar to that in doped materials. These linewidths will also include contributions from defects already present in the crystal, and may have scope for further improvement using optimized substrates. The presence of native \er{} ions in our unimplanted \pbwo{} samples enables a direct comparison of inhomogeneous linewidths between doped and implanted samples; here, we observe that the implanted sample is approximately three times broader than the doped \er{}, but with a linewidth sufficiently narrow to resolve several hyperfine-related features. 

\begin{table}[]
\begin{tabular}{|c|c|}
\hline
Material & Inhomogeneous Linewidth \\ \hline
MgO      & $<$2\,GHz        \\ \hline
\ptosto{}      & 9\,GHz \\ \hline
ZnS      & $>$25\,GHz \\ \hline
\pbwo{}      & 1\,GHz \\ \hline
\tio{}      & 500\,MHz \\ \hline
\moo{}      & 90\,GHz \\ \hline
ZnO      & 1.5\,GHz \\ \hline

\end{tabular}
\caption{Summary of the observed inhomogeneous linewidths for \zy{1}{1} transitions after thermal annealing.}
\label{Tab:linewidths}
\end{table}

A necessary tool in realizing these GHz-wide inhomogeneous linewidths, however, is thermal processing. The annealing conditions for each sample are given in the appropriate material section, but the effects we observe can be generally categorized as changing site occupancies and changing inhomogeneous linewidths. These effects are summarized in Table \ref{Tab:anneal}. The effect of annealing is particularly striking in \pbwo{} and ZnO; in these cases, annealing significantly changes the nature of the implanted \er{} ions, which are not initially optically active in the case of \pbwo{} and occupy a highly disordered site in the case of ZnO. The thermal processing conditions we outline for each material do not necessarily represent the optimal conditions; given the complex interplay of \er{} diffusion, charge compensation, and implantation damage repair there is likely scope for improvement for each of the materials outlined here.

\begin{table}
\begin{tabular}{|p{0.2\linewidth}||p{0.35\linewidth}||p{0.35\linewidth}|}
\hline
 Material &  \multicolumn{2}{c|}{Effect of Annealing} \\ \hhline{|=|==|}
\hline
  & Site Occupancy & Inhomogeneous Linewidth \\ \hhline{|=|=|=|}
MgO & $O_h$ site increases, additional sites also appear & Narrows, quartet splitting decreases \\ \hline
ZnS & Site 2 appears & No change  \\ \hline
\pbwo{} & No implanted \er{} observable prior to annealing; large increase after anneal & N/A  \\ \hline
\tio{} & Increase in $D_{2d}$ site & Narrows\\ \hline
\moo{} & Minor peaks disappear on annealing & No change  \\ \hline
ZnO & Initial site disappears, replaced by new site & New site has tenfold narrower inhomogeneous linewidth \\\hline
\end{tabular}
\caption{Summary of the effects of annealing on different host materials}
\label{Tab:anneal}
\end{table}

\section{Outlook and Conclusions}
Several materials presented here are promising candidates for future single-\er{} studies. MgO is a low refractive index material with reports of millisecond spin $T_1$ for \er{} at temperatures $>$ 1\,K\cite{Borg1970}, while \pbwo{} and \cawo{} have interesting hyperfine structure which may be useful for initializing ancilla qubits. Beyond simply replacing yttrium-based crystals as passive host materials, both \ptosto{} and ZnO (which are respectively ferroelectric and piezoelectric) have promising optical properties. The prospect of long optical and spin coherence in polar crystals is particularly intriguing; the ability to actively control the host material could enable rapid frequency tuning of optical transitions, novel control schemes\cite{George2013,Liu2021}, and spectral shaping of emitted photons\cite{Lukin2020}. 

While our discussion has focused on integration with nanophotonic devices, many of these materials may be useful in ensemble applications. Our results show that ion implantation can be used to identify promising candidate materials, which could then be doped during crystal growth. This would provide a route to combining the flexibility and relative ease of sample preparation associated with ion implantation with the higher doping levels and high-quality materials possible with careful crystal growth. More generally, the combination of ion implantation and thermal processing we present will enable efficient searches through material parameter space, providing new data which could be used in future computational or data-mining studies to identify new candidate hosts\cite{Ferrenti2020}. This approach is not limited to screening host materials, but could also be used to efficiently search for new emitters in established host materials\cite{Thiel2011,Wolfowicz2021,Graham2017a}.

\section{Acknowledgements}
Funding for this research was provided by the AFOSR (contract FA9550-18-1-0334), the Eric and Wendy Schmidt Transformative Technology Fund at Princeton University, and the Princeton Catalysis Initiative. The materials processing work was supported by the U.S. Department of Energy, Office of Science, National Quantum Information Science Research Centers, Co-design Center for Quantum Advantage (C2QA) under contract number DE-SC0012704. P.G. acknowledges funding from ANR Mirespin project, grant 
ANR-19-CE47- 0011 of the French Agence Nationale de la Recherche. The work at UC Berkeley was supported by the Center for Probabilistic Spin
Logic for Low-Energy Boolean and Non-Boolean Computing (CAPSL), one of the Nanoelectronic
Computing Research (nCORE) Centers as task 2759.002, and a Semiconductor Research
Corporation (SRC) program sponsored by the NSF through CCF 1739635

\bibliography{ErHosts}

\end{document}


\title{Supplementary Information for Erbium-Implanted Materials for Quantum Communication Applications}

\author{Paul Stevenson}
\affiliation{Department of Physics, Northeastern University, Boston, Massachusetts 02115, USA}

\author{Christopher M Phenicie}
\affiliation{Department of Electrical Engineering, Princeton University, Princeton, New Jersey 08544, USA}

\author{Isaiah Gray}
\affiliation{Department of Electrical Engineering, Princeton University, Princeton, New Jersey 08544, USA}

\author{Sebastian P Horvath}
\affiliation{Department of Electrical Engineering, Princeton University, Princeton, New Jersey 08544, USA}

\author{Sacha Welinski}
\affiliation{Thales Research and Technology, 1 Avenue Augustin Fresnel, 91767 Palaiseau, France}

\author{Austin M Ferrenti}
\affiliation{Department of Chemistry, Princeton University, Princeton, New Jersey 08544, USA}

\author{Alban Ferrier}
\affiliation{Chimie ParisTech, PSL University, CNRS, Institut de Recherche de Chimie Paris, Paris, France, 75005}
\affiliation{Faculté des Sciences et Ingénierie,  Sorbonne Université, UFR 933, 75005 Paris, France}

\author{Philippe Goldner}
\affiliation{Chimie ParisTech, PSL University, CNRS, Institut de Recherche de Chimie Paris, Paris, France, 75005}

\author{Sujit Das}
\affiliation{Department of Materials Science and Engineering, University of California, Berkeley, Berkeley, 94720, California, USA}
\affiliation{Department of Physics, University of California, Berkeley, 94720, California, USA}

\author{Ramamoorthy Ramesh}
\affiliation{Department of Materials Science and Engineering, University of California, Berkeley, Berkeley, 94720, California, USA}
\affiliation{Materials Sciences Division, Lawrence Berkeley National Laboratory, Berkeley, 94720, California, USA}
\affiliation{Department of Physics, University of California, Berkeley, 94720, California, USA}

\author{Robert J Cava}
\affiliation{Department of Chemistry, Princeton University, Princeton, New Jersey 08544, USA}

\author{Nathalie P de Leon}
\affiliation{Department of Electrical Engineering, Princeton University, Princeton, New Jersey 08544, USA}

\author{Jeff D Thompson}
\email{jdthompson@princeton.edu}
\affiliation{Department of Electrical Engineering, Princeton University, Princeton, New Jersey 08544, USA}

\maketitle

\onecolumngrid
\section{Experimental Methods and Analysis}
\subsection{Resonant Fluorescence Measurements}
The millisecond radiative lifetime of \er enables resonant fluorescence experiments to be performed using mechanical gating elements. A tunable diode laser (Toptica CTL 1500) pumps an erbium-doped fiber amplifier to provide the excitation light. This excitation is modulated by a mechanical chopper (Thorlabs MC2000B) with a 10\% duty cycle. Emission is collected from the sample when the excitation light is blocked by using a second chopper out of phase with the excitation chopper. A polarizing beam-splitter is used to further filter residual excitation light.

The emitted light is detected either by an amplified InGaAs photodiode (Femto systems OE-200-IN2) and a lock-in amplifier (SRS SR860) in the case of the excitation spectra, or by a spectrometer and cooled InGaAs diode array (Princeton Instruments HRS-300 + PyLon detector) to resolve the emission spectrum. The excitation wavelength was monitored with a wavemeter (Bristol Instruments).

\subsection{Sample Preparation and Thermal Processing}
All samples were implanted with \er{} at a fluence of $10^{12}$ cm$^{-2}$, with the recipe detailed in Table \ref{Table:recipe}, except \ptosto{} which was doped during growth of the Pb-containing layer. 

All samples were initially studied as-implanted \textit{i.e.} without annealing. Samples were subsequently annealed to repair implantation damage, under conditions detailed in each section. Annealing protocols were drawn from available literature on the material and are detailed in each material section.

\begin{table}[h]
\begin{tabular}{|c|c|c|}
\hline
Step & Fluence /cm$^{-2}$ & Energy / keV \\ \hline
1 & $3.75\times 10^{12}$ & 10 \\ \hline
2 & $5\times 10^{12}$ & 25 \\ \hline
3 & $7.5\times 10^{12}$ & 50 \\ \hline
4 & $1\times 10^{13}$ & 100 \\ \hline
5 & $1.25\times 10^{13}$ & 150 \\ \hline
6 & $1.25\times 10^{13}$ & 250 \\ \hline
7 & $3.75\times 10^{13}$ & 350 \\ \hline
\end{tabular}
\caption{\label{Table:recipe} Implantation recipe used to generate \er{}-implanted samples, unless otherwise stated in the main text.}
\end{table}
Samples were annealed under either an air atmosphere or under vacuum in a tube furnace.

\subsection{\ptosto{} Sample Preparation}
Pb$_{1.199}$Er$_{0.001}$TiO$_3$ films with a thickness of 100 nm were synthesized on \tio{}-terminated single-crystalline SrTiO$_3$ (001) substrates via reflection high-energy electron diffraction (RHEED)-assisted pulsed-laser deposition (KrF laser). The Pb$_{1.199}$Er$_{0.001}$TiO$_3$ layers were grown at 580\,$^{\degree}$C in 100\,mTorr oxygen pressure. The laser fluence was 1.5 J/cm$^2$ with a repetition rate of 10 Hz. RHEED was used during the deposition to ensure the maintenance of a layer-by-layer growth mode for the Pb$_{1.199}$Er$_{0.001}$TiO$_3$3. After deposition, the thin films were annealed for 10 minutes in 50\,Torr oxygen pressure to promote full oxidation and then cooled down to room temperature at that oxygen pressure.

\subsection{Analysis of Excitation-Emission Spectra}
The central assumption in our analysis is that thermalization in the excited state is fast relative to the (millisecond) excited state lifetime, such that all excitations associated with a particular \er site have the same emission spectrum.

Assigning $Z$ and $Y$ levels begins with identifying the \zy{1}{1} transition. When thermal occupation of excited crystal field states is negligible, this is the lowest energy excitation peak and the highest energy emission peak. With \zy{1}{1} assigned, the excitation and emission spectra can be used to generate initial guesses for the $Z_n$ and $Y_n$ levels. These assignments are checked for self-consistency (\textit{e.g.} that thermally occupied states predicted by the splittings are indeed observed), and if required, refined and validated using spectra at higher temperatures to thermally populate other crystal-field levels.

\section{Additional Discussion}

\subsection{Emission from Thermally Occupied States in MgO}
\begin{figure}[]
  \centering
  \includegraphics{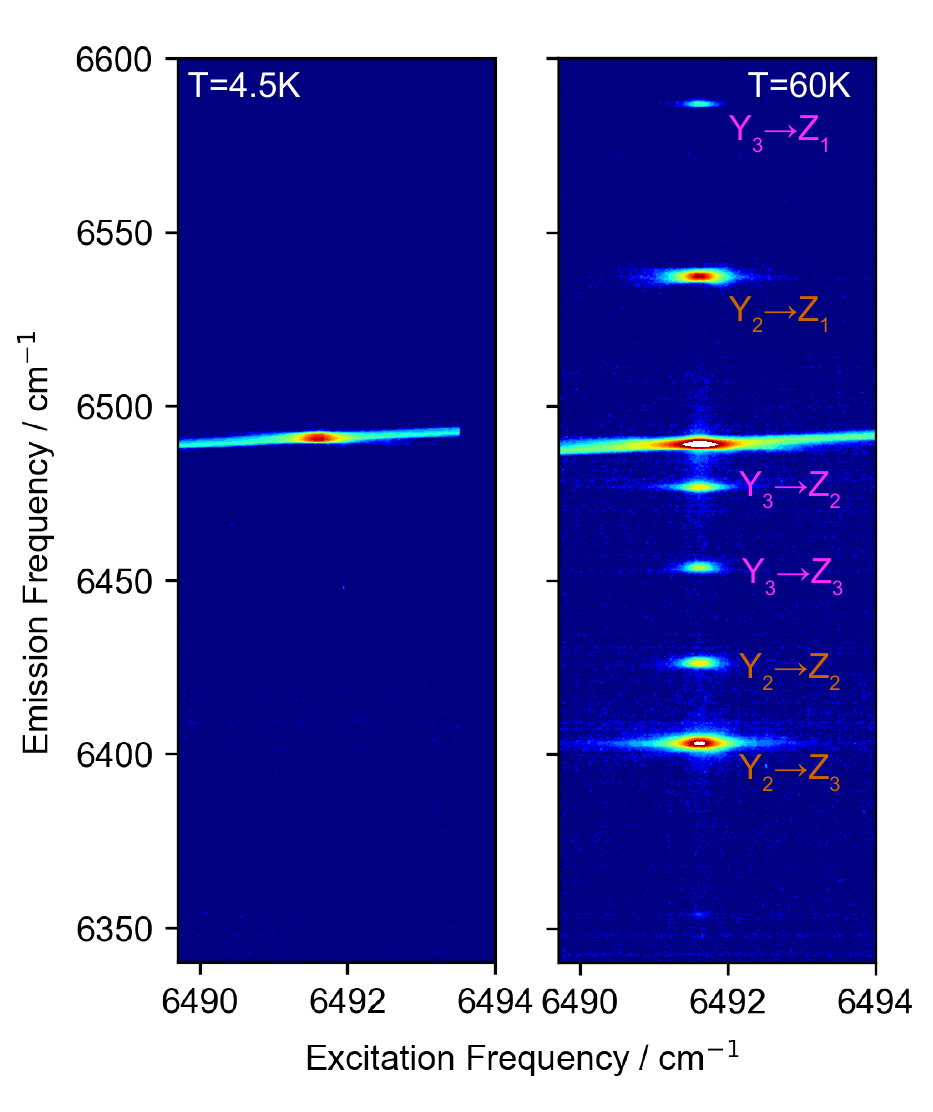}
  \caption{Emission spectrum for the $O_h$ site in MgO at 4.5\,K and 60\,K. At elevated temperature, additional peaks are observed which originate from thermally-occupied excited states.} 
  \label{Fig:SI_MgO_thermal}
\end{figure}

As noted in the main text, emission from the $Y_1$ level to certain $Z$ levels is symmetry-forbidden for the $O_h$ site. To determine the relative positions in the ground state, we measure the emission spectrum for this site at 60\,K, where emission from the thermally populated $Y_2$ and $Y_3$ sites is observed. This data is shown in Figure \ref{Fig:SI_MgO_thermal}.

\subsection{Other \er{} Sites in MgO}
\er{} implanted MgO presents a particularly rich site-selective spectrum. Our analysis in the main text focuses on one site, which we identify as the cubic substitutional site, but our data allows us to tabulate all the sites we observe. These sites likely arise from lower symmetry sites associated with different environments of the \er{} (\textit{e.g.} interstitial, nearby damage centers, and combinations thereof). The detailed assignment of these sites is beyond the scope of this work, but we present the site energies we determine here for completeness. The experimentally-determined energy levels are given in Table \ref{Tab:SI_MgO}. Note that most of these sites are significantly less populated than the primary octahedral site discussed in the main text; only sparse information on energy levels is available in some cases. 

\begin{figure}[]
  \centering
  \includegraphics{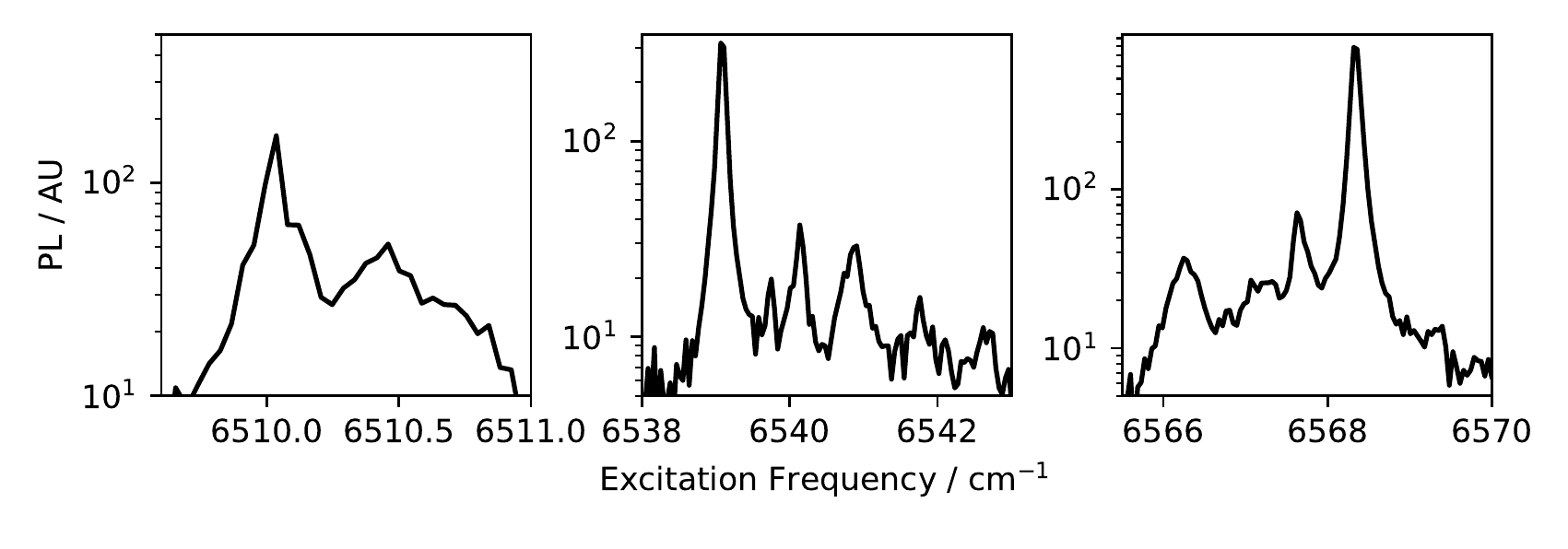}
  \caption{Excitation spectra of the Site 2 in MgO. The closely spaced clusters of peaks are consistent with a local charge-compensation environment. Note that the separation of these peaks is on the order of 1\wn{} or smaller.} 
  \label{Fig:SI_MgO_CC}
\end{figure}

One of these minor sites (``Site 2'') shows a number of clustered peaks around each peak, shown in Figure \ref{Fig:SI_MgO_CC}. These peaks exceed the number of transitions which can arise from the $Z$ and $Y$ manifolds of a single site. The small ($\approx 1$\wn{}) splittings around each peak is suggestive of minor perturbations of a single site; similar features were observed for Nd$^{3+}$ in CaF$_2$ and SrF$_2$ and were assigned the variations in the position of local charge compensation structural features.

\begin{table}[h!]
\begin{tabular}{|c||c|c||c|c||c|c||c|c|}
\hline
 & \multicolumn{2}{c||}{Site 2} & \multicolumn{2}{c||}{Site 3} & \multicolumn{2}{c||}{Site 4} &\multicolumn{2}{c|}{Site 5} \\ \hhline{|=|==|==|==|==|}
 \zy{1}{1} / \wn{} & \multicolumn{2}{c||}{6510.0 } & \multicolumn{2}{c||}{6499.1} & \multicolumn{2}{c||}{6557.0 } & \multicolumn{2}{c|}{6534.1} \\ \hline
$n$ & Z$_n$ & Y$_n$ & Z$_n$ & Y$_n$ & Z$_n$ & Y$_n$ & Z$_n$ & Y$_n$\\ \hline
1 & 0.0 & 0.0 & 0.0 & 0.0 & 0.0 & 0.0 & 0.0 & 0.0 \\ \hline
2 & 41.0 & 29.1 & 27.0 & 34.9 & 82.8 & 42.3 & 81.0 & 46.4 \\ \hline
3 & 83.6 & 58.4 & 96.0 & 71.8 & * & * & 114.0 & * \\ \hline
4 & 97.2 & * & 128.6 & * & * & * & 151.3 & * \\ \hline
5 & 156.0 & * & * & * & * & * & * & * \\ \hline
6 & 187.3 & * & * & * & * & * & * & * \\ \hline
\end{tabular}
\caption{Experimentally determined energy levels for other sites in MgO. The nature of these sites is undetermined; however, Site 2 site has a cluster of peaks suggestive of local charge compensation.}
\label{Tab:SI_MgO}
\end{table}

\subsection{Annealing behavior of ZnO}
\begin{figure}[]
  \centering
  \includegraphics{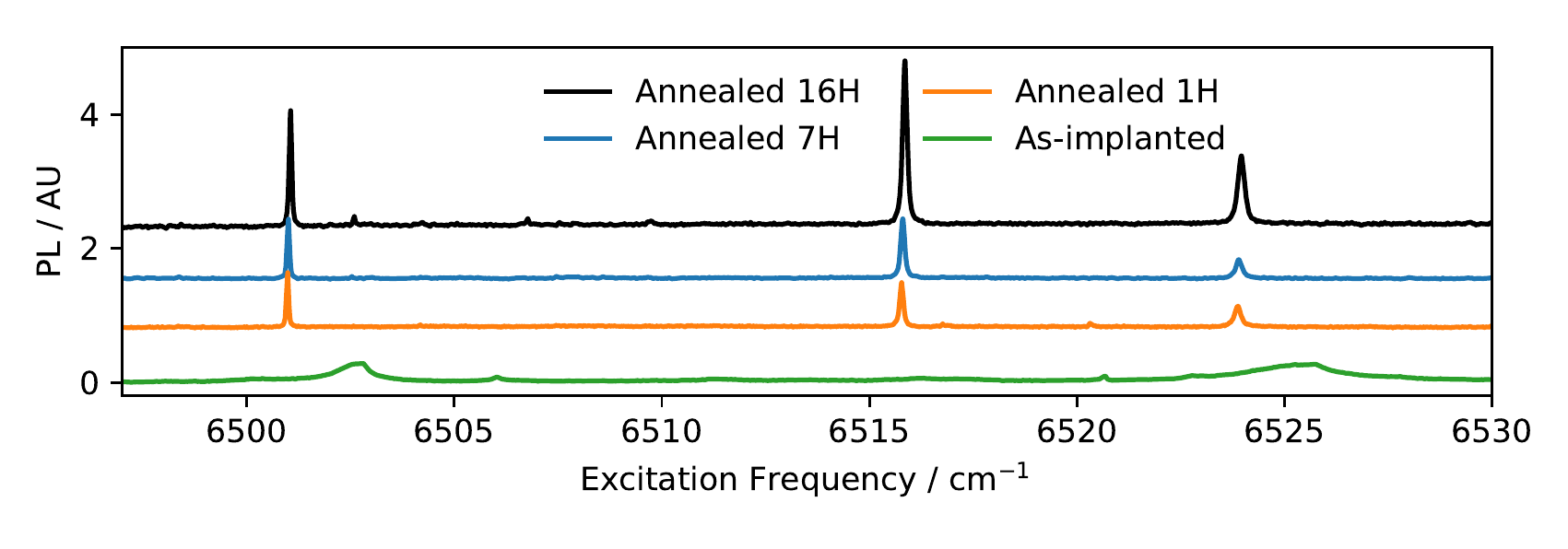}
  \caption{Comparison of excitation spectra for as-implanted and various annealing times ZnO, showing the substantial inhomogeneous narrower after annealing. A new set of peaks with substantially narrower inhomogeneous linewidths appears after annealing.} 
  \label{Fig:SI_ZnO}
\end{figure}

As-implanted ZnO shows many broad ($>10$\,GHz) peaks in the excitation spectrum. Upon annealing at 750\dc{} in air for a total of 16 hours, many of these peaks disappear, as shown in Figure \ref{Fig:SI_ZnO}. The peaks which remain are substantially narrower (\zy{1}{1} inhomogeneous linewidth 1.5\,GHz at 4\,K) and are consistent with a single environment of \er{} (see main text for excitation-emission spectrum).

\section{Excited State Lifetime}
\begin{figure}[h!]
  \centering
  \includegraphics{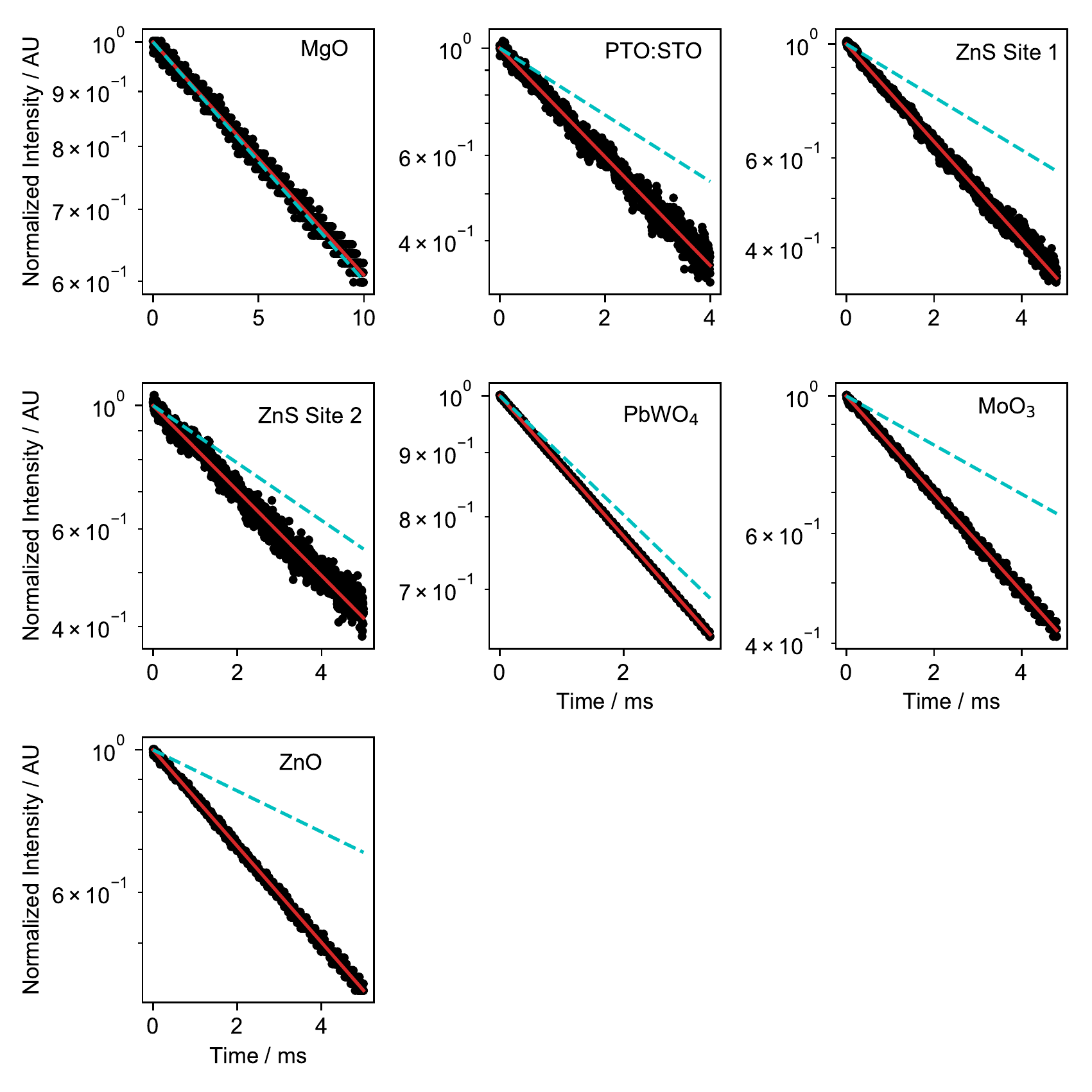}
  \caption{Lifetime data (black dots) and fits (red lines). The expected decay from a pure magnetic dipole transition is shown in blue. Parameters are given in Table \ref{tab:SI_lifetimefits}} 
  \label{Fig:SI_lifetime}
\end{figure}

The data and fits for the excited state lifetime measurements on each sample are shown in Figure \ref{Fig:SI_lifetime}. The $>$ms lifetime expected for the \er{} magnetic dipole allows this to be measured in a straightforward manner by mechanically chopping the excitation and recording the response of a photodiode. Uncertainties on the lifetime are given as $\pm1\sigma$ for each dataset.

For reference, we also reproduce the expected lifetimes for a pure magnetic dipole transition. This can be estimated using the free ion radiative rate, modified by the refractive index scaling of density of states ($n^3$). As discussed in the main text, lifetimes shorter than predicted by a pure MD transition can occur either because of competing non-radiative pathways, or because of mixing between $d$ and $f$ orbitals allowing some electric dipole character. The lifetime measurement cannot distinguish between these two scenarios; however, we note that in the two systems where we expect no $d$-$f$ mixing (MgO, \tio{}) we do indeed see MD-limited lifetimes, which suggests that ion implantation damage does not necessarily result in decreased radiative yield.

\begin{table}[h!]
\begin{tabular}{|c|c|c|}
\hline
Host & Excited State Lifetime / ms & Magnetic Dipole-Limited Lifetime / ms \\ \hhline{|=|=|=|}
MgO & 20.1 (0.03) & 19.6 \\ \hline
\ptosto{} & 3.9 (0.01) & 6.3 \\ \hline
ZnS (Site 1) & 4.6 (0.01) & 8.4 \\ \hline
ZnS (Site 2) & 5.7 (0.02) & 8.4 \\ \hline
\pbwo{} & 7.7 (0.05) & 9.1 \\ \hline
\moo{} & 5.5 (0.04) & 11.0 \\ \hline
ZnO & 5.8 (0.04) & 13.6 \\ \hline
\end{tabular}
\caption{Fit parameters and uncertainties for the lifetime measurements described in the main text and shown in Figure \ref{Fig:SI_lifetime}. Uncertainties are given in parenthesis.}
\label{tab:SI_lifetimefits}
\end{table}

\clearpage